\documentclass[superscriptaddress,amsmath,amssymb, aps, prl,twocolumn]{revtex4-2}

\usepackage{graphicx}
\usepackage{dcolumn}
\usepackage{bm}
\usepackage[colorlinks]{hyperref}
\usepackage{cancel}
\usepackage{amsthm}

\newcommand{\be}{\begin{equation}}
\newcommand{\ee}{\end{equation}}
\newcommand{\beq}{\begin{eqnarray}}
\newcommand{\eeq}{\end{eqnarray}}

\newcommand{\ket}[1]{\mbox{$ | #1 \rangle $}}
\newcommand{\bra}[1]{\mbox{$ \langle #1 | $}}

\newcommand{\expval}[1]{\mbox{$\langle #1 \rangle$}}

\newcommand{\orcid}[1]{\href{https://orcid.org/#1}{\includegraphics[width=7pt]{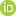}}}

\begin{document}

\title{Experimental investigation of a quantum heat engine powered by generalized measurements}

\author{V. F. Lisboa}
\affiliation{Center for Natural and Human Sciences, Federal University of ABC, Avenida dos Estados 5001, 09210-580, Santo Andr\'{e}, S\~{a}o Paulo, Brazil}

\author{P. R. Dieguez\orcid{0000-0002-8286-2645}}\email{dieguez.pr@gmail.com}
\affiliation{International Centre for Theory of Quantum Technologies (ICTQT), University of Gdańsk, Jana Bazynskiego 8, 80-309 Gdańsk, Poland}
\affiliation{Center for Natural and Human Sciences, Federal University of ABC, Avenida dos Estados 5001, 09210-580, Santo Andr\'{e}, S\~{a}o Paulo, Brazil}

\author{J. R. Guimar\~{a}es\orcid{0000-0001-5649-5557}}
\affiliation{Center for Natural and Human Sciences, Federal University of ABC, Avenida dos Estados 5001, 09210-580, Santo Andr\'{e}, S\~{a}o Paulo, Brazil}
\affiliation{Peter Grünberg Institute, Institute for Functional Quantum Systems (PGI-13), Forschungszentrum Jülich, North Rhine-Westphalia, 52425 Jülich, Germany.}

\author{J. F. G. Santos} 
\affiliation{Center for Natural and Human Sciences, Federal University of ABC, Avenida dos Estados 5001, 09210-580, Santo Andr\'{e}, S\~{a}o Paulo, Brazil}
\affiliation{Division of Natural and Applied Sciences, Duke Kunshan University, Kunshan, Jiangsu, 215300, China}

\author{R. M. Serra\orcid{0000-0001-9490-3697}}\email{serra@ufabc.edu.br}
\affiliation{Center for Natural and Human Sciences, Federal University of ABC, Avenida dos Estados 5001, 09210-580, Santo Andr\'{e}, S\~{a}o Paulo, Brazil}
\affiliation{Department of Physics, Zhejiang Normal University, Jinhua 321004, China}

\begin{abstract}
Generalized measurements may allow the control of its back-action on the quantum system by interpolating from a very weak to strong projective action. Such a measurement can fuel a quantum heat engine or extract work depending on the system-meter interaction. Here, we performed a proof-of-concept experiment using nuclear magnetic resonance techniques to investigate a spin quantum heat engine driven by non-selective generalized (weak) measurements without feedback control. Our prototype of a quantum thermal device operates with a measurement protocol and a single heat bath. The protocol is composed of two non-selective measurement channels with variable measurement strengths, one dedicated to fueling the device (analogous to a hot heat source) and the other committed to work extraction. The experimental results highlight that this kind of quantum thermal device can reach unit efficiency with maximum extracted power by fine-tuning of the measurement strengths. 
\end{abstract}

\maketitle

\section{Introduction}

Heat engines were crucial in the developments of modern societies and for classical thermodynamics, providing fundamental constraints on physically allowed processes and the understanding of how to convert stochastic energy gained in the form of a heat transfer, into useful work~\cite{Ce2001book}. 
The recent downscale of thermal devices led to
the development of quantum versions of heat engines~\cite{Myers2022,campisi2011colloquium,campisi14,batalhao14,elouard15,campisi15,dechant15,campisi16, Peterson2018, Klatzow2019, Zagoskin2012, Altintas2015, Campisi2016, Rossnagel2016, Barontini2019, Bouton2021}. Thermal and quantum fluctuations~\cite{esposito2009nonequilibrium,quan07} become
relevant and must be taken into account for a proper understanding of the
thermodynamics in such micro and nano-scales
\cite{Alicki2019,Vinjanampathy2016,Kosloff2013}. Furthermore, quantum thermal
devices are employed theoretically and experimentally, to investigate thermodynamic uncertainty relations \cite{Timpanaro2019,Lee2021,Sacchi2021}, fluctuation theorems \cite{batalhao14,berg2018,micadei2020, campisi2011colloquium, campisi14,Denzler2021}, and other subjects~\cite{Myers2022}.

A figure of merit used to
quantify the performance of heat engines is the efficiency, defined
as the ratio between the extracted work and the heat absorbed from a hot source. The best efficiency is reached for a reversible
cycle but the price, in general, is zero or very small output power. According to the Carnot limit, it is restricted by $\eta_{\text{Carnot}}=1-T_{c}/T_{h}$,
with $T_{c,h}$ denoting the temperature of the cold and hot heat baths, respectively. For quantum heat engines operating in a finite-time regime, the performance depends directly on the irreversibility of the processes, which can be quantified by the entropy production \cite{Groot1984,Lebon2008,batalhao15,camati16}. Quantum features such as coherence, for example, can influence the performance of quantum thermal devices improving or deteriorating it \cite{Henao2018,Micadei2019,Camati2019,Santos2019,Francica2019}.

Apart from heat engines operating between two thermal reservoirs, Szilard \cite{Szilard1929,Quan2006}
proposed long ago a gedanken experiment where work is extracted from a single
heat bath through a feedback mechanism performed by a Maxwell's demon \cite{Leff1990}. Although the idea of
feedback control goes back to the beginning of the classical thermodynamics development, it has recently become even more relevant in designs of small quantum devices \cite{camati16,Elouard2017,Elouard2018},
being useful mostly to control irreversibility and to optimize work extraction. Measurement-powered engines use the fact that measurements on a quantum
system may change its state and internal energy  \cite{Talkner2017,Elouard2017-1,Brandner2015, Lin2021, Su2021}. Employing non-selective measurements, it was recently theoretically proposed a single temperature heat engine without feedback control~\cite{Talkner2017}, where a
standard thermal reservoir that provides energy to an Otto cycle is replaced by a measurement device. One important step further
was given by introducing the concept of quantum heat  \cite{Elouard2017},
identified with the stochastic energy fluctuations taking place during a quantum measurement. Moreover, several measurement-based thermodynamic protocols have been studied in the last few years~\cite{Bresque2021,Campisi2019,Ding2018,Jordan2020,Mohammady2017,Campisi2017,chand2017single,chand2017measurement,chand2018critical,Elouard2017,Elouard2018,anka2021measurement,alam2022two}.

 Considering the development of quantum heat engines powered by projective measurements, it is natural to ask under what assumptions it is possible to perform a heat engine cycle by using generalized measurements described by positive operator-valued measurements (POVMs), since this approach includes phenomena not completely captured by projective measurements, such as detectors with non-unit efficiency, measurement outcomes with additional randomness, and in particular, weak measurements which have found numerous applications in quantum thermodynamics tasks~\cite{jacobs09,alonso16,pati2020,mancino18}.  Since any generalized measurement can be implemented as a sequence of weak measurements~\cite{oreshkov05,Dieguez18}, this also implies that the intensity of generalized measurements can be continuously adjusted to interpolate between the weak and strong regimes~\cite{pan16,mancino18}. Weakly measured systems are relevant for consistent observations of work and heat contributions to an externally driven quantum stochastic evolution~\cite{alonso16}. From the experimental point of view, work and heat associated with unitary and nonunitary dynamics along single quantum trajectories, as well as  information dynamics of a quantum Maxwell's demon, were investigated by employing superconducting qubits weakly coupled to a meter system~\cite{naghiloo2020heat,naghiloo2018information}. Also, a new class of Maxwell's demon was realized with a tunable dissipative strength using a nitrogen-vacancy center~\cite{hernandez2022autonomous}. %

Interestingly, it is possible to design a deterministic protocol employing generalized measurements with variable strength to perform a heat engine cycle~\cite{behzadi2020quantum}.
Here, we report an experimental investigation of a  quantum heat engine powered by generalized measurements inspired in Ref. \cite{behzadi2020quantum}. Employing a liquid state nuclear magnetic resonance (NMR) setup \cite{oliveira2007}, we emulate two generalized measurement channels to implement a proof-of-concept three stroke engine cycle (two measurements and one heat sink). The experimental results highlight that this kind of quantum thermal device can reach unit efficiency with maximum extracted power. 

\section{Experimental Investigation}

In the reported experiment, we used a liquid sample of $^{13}$C-labeled sodium formate (HCO$_2$Na) diluted in deuterium oxide (D$_2$O). Each molecule in the sample is composed of four different nuclei (isotopes): $^{1}$H, $^{13}$C, $^{18}$O, and $^{23}$Na. To look over a generalized measurement-powered cycle, the experimental control was focused on the $^{1}$H and $^{13}$C nuclei which have nuclear spins $1/2$. This sample was prepared with a low dilution ratio ($\approx2\%$), such that the inter-molecular interactions can be neglected and the sample can be considered as a set of almost identically prepared pairs of two spin-$1/2$ systems. The $^{18}$O has nuclear spin $0$, $^{23}$Na has nuclear spin $3/2$, and both nuclei will not play an important role in the experiment. The deuterium in the solvent is used for locking the resonance of the longitudinal magnetic field in the setup.

The experiment was performed using a Varian 500 MHz spectrometer equipped with a double-resonance probe head with a magnetic field-gradient coil. The sample is inserted inside a superconducting magnet that produces a longitudinal ($z$-axis) static field with intensity, $B_{0}\approx 11.75$~T. Under the action of this static field, the resonance frequencies of $^{1}$H and $^{13}$C nuclei are approximately $500$ and $125$~MHz, respectively. The states of the nuclear spins were controlled using electromagnetic field pulses with time-modulated radio frequencies (rfs) applied in the transverse direction ($x$ and $y$-axes), longitudinal magnetic field gradients, as well as by sequences of free evolution of the system under the action of the scalar coupling. The latter is described by the Hamiltonian $H_J=\frac{h}{4}J\sigma_z^{\text{H}}\otimes \sigma_z^{\text{C}}$, where $J \approx 194.65$~Hz is the coupling constant between the $^{1}$H and $^{13}$C.

\begin{figure}[h]
\centering
\includegraphics[width=8.50cm]{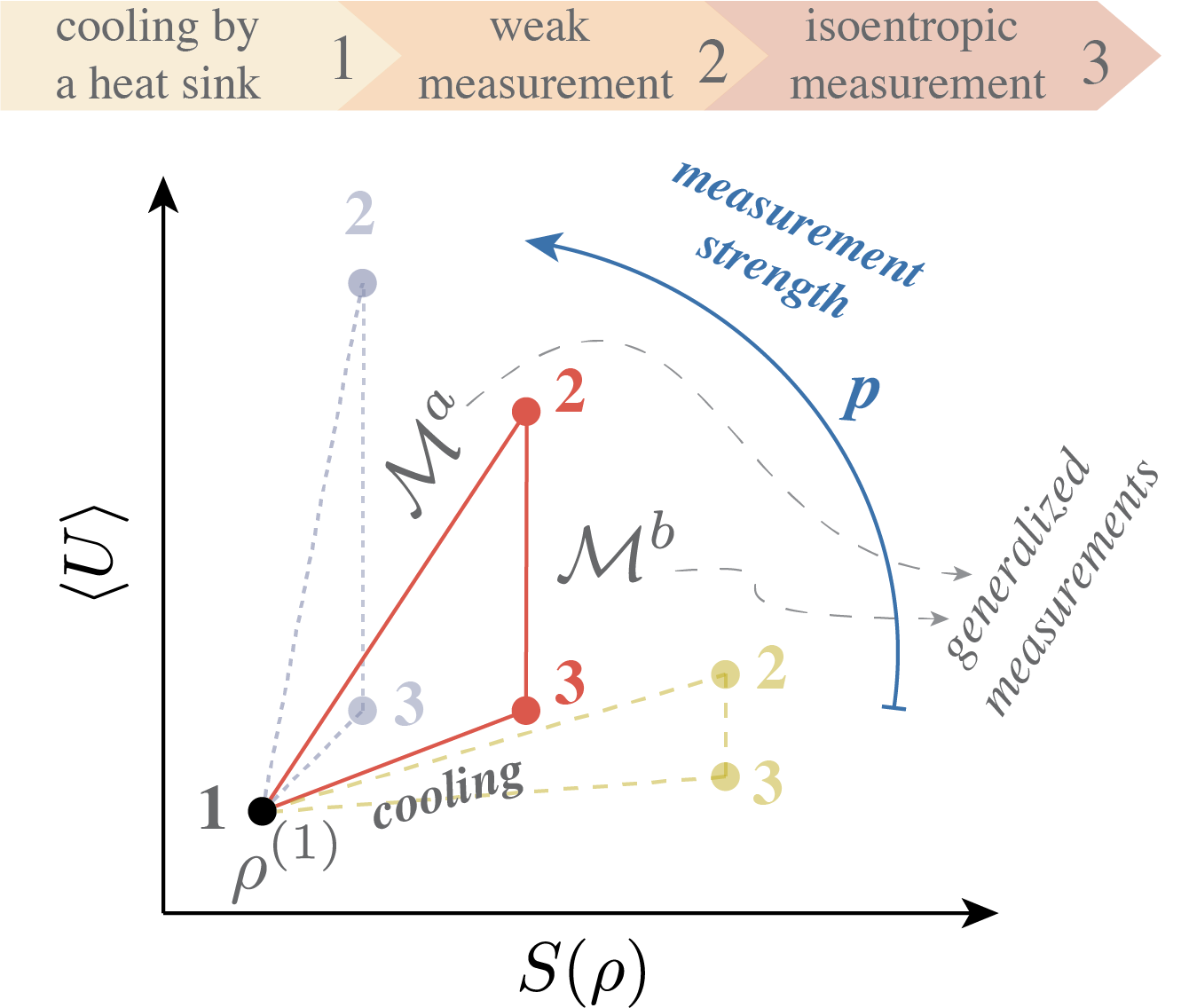}
\caption{Schematic description of the theoretical heat engine cycle powered by generalized measurements. First, the work substance interacts with a cold sink and it is cooled to the equilibrium state $\rho^{(1)}$. Second, a general measurement channel $\mathcal{M}^{a}$ is performed which leads to an increase in the average energy and von Neumann entropy of the working substance. Over this channel, the working substance absorbs energy in a stochastic way (heat) from the meter. Third, another general measurement channel $\mathcal{M}^{b}$ chosen properly to produce an isentropic decrease of the average energy of the working substance, is applied. The energy exchanged with the meter in the latter channel has a work nature. The scheme also illustrates the variation of the internal energy and entropy for different measurement strengths.} 
\label{Fig1}
\end{figure}
The $^{13}$C nucleus will play the role of working substance while the $^{1}$H nucleus acts as an ancillary system for the implementation of the measurement channels. The rf fields on resonance or near to resonance with $^{13}$C will take the place of a cold heat sink.

\begin{figure*}
\centering
\includegraphics[width=0.85\textwidth]{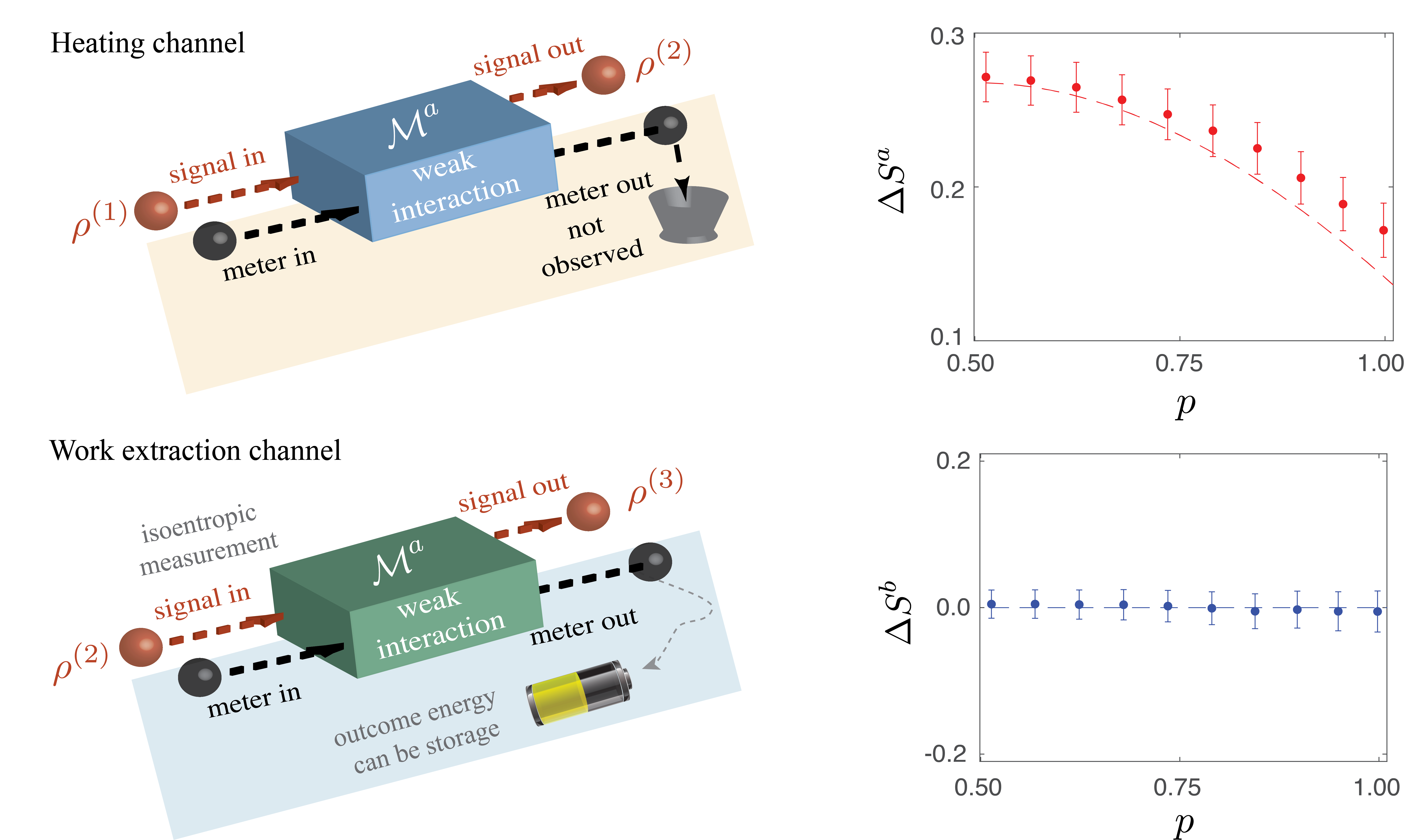}
\caption{Sketch of the two generalized measurement channels implementations and the corresponding entropy variations. In the measurement channel $\mathcal{M}^a$ (heating channel), the ancillary quantum degree of freedom of the meter interacts with the working substance with controlled strength. Such meter's degrees of freedom will not be observed leading the working substance to the state $\rho^{(2)}=\mathcal{M}^a(\rho^{(1)})$. This channel acts as a heat source. In the second measurement channel $\mathcal{M}^b$ (work extraction channel) the system interacts again with the internal degree of freedom of the meter, work is extracted, and could, in principle, be stored for further use. This channel is adjusted to produce an isentropic variation of the system's internal energy.  On the right side, the von Neumann entropy change, $\Delta S^{a}$ and $\Delta S^{b}$, after the channels $\mathcal{M}^a$ and $\mathcal{M}^b$, are displayed respectively, as a function of the measurement strength. The dashed curves are numerical simulations of the channels experimental implementation. The symbols are experimental data for the sequential application of the channels with an initial state at a spin temperature equivalent to $k_BT=2.98\pm0.19$~peV.} 
\label{Fig2}
\end{figure*}

In general, we can model a measurement apparatus as an object that has some microscopic quantum degrees of freedom (that interacts with the system to be measured) which is further amplified to the classical realm through  a macroscopic pointer that is coupled with the former. As we are not interested in a specific outcome of the measurement, we can model such a non-selective measurement by simply discarding (tracing over) the quantum degrees of freedom of the meter. Then, the result of this procedure is the average statistics over all the possible outcomes of the measurement. For our practical purposes, the generalized measurement will be implemented by the interaction of the working substance with one non-observed auxiliary system, resulting in a completely positive trace-preserving map acting on the working substance. 

The generalized measurement powered engine cycle consists
of three strokes, depicted in Fig. \ref{Fig1}, which are described as follows.


Stroke one: Initially the $^{13}$C nuclear spins are cooled. This
is effectively performed by employing spatial average techniques \cite{batalhao14,batalhao15,camati16,Micadei2019,micadei2020},
resulting in a state equivalent to the Gibbs state $\rho^{(1)}=\text{exp}\left[-\beta H^{\text{C}}\right]/\mathcal{Z}$ (see Appendix A),
at cold inverse spin temperature $\beta=1/\left(k_{B}T\right)$, where
$\mathcal{Z}$ is the partition function. The system's  Hamiltonian is associated to an offset of the transverse rf fields with relation to the resonance of the $^{13}$C nucleus, given by $H^{\text{C}}=-\left(h\nu/2\right)\sigma_{z}$ \cite{Micadei2021}, with $\sigma_{\ell}$, $\ell=\left(x,y,z\right)$
standing for the Pauli matrices and we have adjusted the frequency as $\nu \approx 1$~kHz.

Stroke two: The first generalized non-selective measurement is performed on the working
substance ($^{13}$C nucleus), which is effectively implemented by a controlled interaction with an ancillary system ($^{1}$H nucleus), prepared in an initial state equivalent to $\rho^{\text{H},0}=\ket 0 \bra 0$. The ancillary system is not observed and after the interaction it can be traced out as depicted in Fig. \ref{Fig2}. The interaction with the ancillary system is managed to assemble the desired measurement map, leading to,  
$\mathcal{M}^{a}:\rho^{(1)} \rightarrow \rho^{(2)}=\sum_{i}{\mathcal{M}_{i}^{a}}\rho^{(1)}{\mathcal{M}_{i}^{a}}^{\dagger}$
where the Kraus operators of the generalized measurement are ${\mathcal{M}^{a}_{1}}=\sqrt{1-p\Omega}|0\rangle\langle0|+|1\rangle\langle1|$
and ${\mathcal{M}^{a}_{2}}=\sqrt{p\Omega}|1\rangle\langle0|$ (satisfying
$\sum_{i}{\mathcal{M}_{i}^{a}}^{\dagger}{\mathcal{M}_{i}^{a}}=I$), and $\Omega=1-e^{-\beta h\nu}$
is a factor that depends on the system temperature after the stroke one. The parameter $p$ will be related to the measurement strength. 
The associated POVMs are explicitly given by ${\mathcal{M}^{a}_{1}}^{\dagger}{\mathcal{M}^{a}_{1}}=(1-p\Omega)\ket{0}\bra{0}+\ket{1}\bra{1}$ and ${\mathcal{M}^{a}_{2}}^{\dagger}{\mathcal{M}^{a}_{2}}=p\Omega\ket{0}\bra{0}$. Then, it is easy to note that when $p\Omega=0$ there is no measurement at all. Meanwhile, for $p\Omega=1$ the channel map corresponds to a full projection. The post-measurement state in this stroke can be written explicitly as $\rho^{(2)}=\frac{1}{2} I+ \left(\frac{1}{2}-p \right) \tanh{\left(\beta  h \nu/2\right) }\sigma_z$.  

In our implantation, we are interested in the range 
$0<p\Omega<1$ where we can vary from a weak to a moderated measurement strength (as we will discuss bellow). In this stage, the internal energy variation of the working substance can be interpreted as the heat absorbed from the meter, whenever its von Neumann entropy increases, as usually adopted~\cite{behzadi2020quantum}. So, the heat absorbed in the stroke two is simply given by 
$\langle \mathcal{Q}^{p}\rangle=\text{tr}\left[H^{\text{C}}\left(\rho^{(2)}-\rho^{(1)}\right)\right]$, being directly associated to the variation of the nuclear spin magnetization. Using the measurement channel, we can write the heat absorbed as
\begin{equation}
 \expval{\mathcal{Q}^{p}}= h \nu p\tanh \left(\frac{\beta h \nu}{2} \right),
 \label{heat}
\end{equation}
and the variation of the von Neumann entropy of the working substance in this stroke is $\Delta S^{a} = \text{S}(\rho^{(2)})-\text{S}(\rho^{(1)})$, where $\text{S}(\rho)= -\text{tr}(\rho\ln \rho)$ is the von Neumann entropy. The measurement channel $\mathcal{M}^{a}$ will fuel the engine as a heat source, when  $\expval{\mathcal{Q}^{p}}>0$ and $\Delta S^{a}>0$. That happens when we tune the measurement strength as constrained by $\frac{1}{2}<p<1$.

Stroke three: With the aid of the ancillary system, a second non-selective generalized measurement channel, as depicted in Fig. \ref{Fig2}, will be applied on the working substance leading to $\mathcal{M}^{b}:\rho^{(2)}\rightarrow\rho^{(3)}=\sum_{j=1}^{2}\mathcal{M}^{b}_{j}\rho^{(2)}{\mathcal{M}^{b}_{j}}^{\dagger}$
with the Kraus operators $\mathcal{M}^{b}_{1}=|0\rangle\langle0|+\sqrt{1-q}|1\rangle\langle1|$
and $\mathcal{M}^{b}_{2}=\sqrt{q}|0\rangle\langle1|$ (satisfying $\sum_{i}{\mathcal{M}_{i}^{b}}^{\dagger}{\mathcal{M}_{i}^{b}}=I$). It is interesting to note that, besides the present application, the weak measurement operator ${\mathcal{M}_{1}^{b}}$ and its corresponding weak reverse operator ${\mathcal{M}_{1}^{a}}$ can be employed in protocols (with selective measurements) to protect entanglement from decoherence (see, for instance,~\cite{Kim2012}).

By imposing 
\begin{equation}
 q=\frac{\left(2p-1\right)\Omega}{\left(p-1\right)\Omega+1},
 \label{parameter}
\end{equation}
we ensure that the von Neumann entropy of the working substance does not change during the measurement channel $\mathcal{M}^{b}$, $\Delta S^{b} = \text{S}(\rho^{(3)})-\text{S}(\rho^{(2)})=0$. In this way, the energy exchanged in the stroke three can be associated to
work delivered by the working substance to the meter \cite{behzadi2020quantum}. The post-measurement state, with the measurement strength parameter adjusted according to Eq. (\ref{parameter}), turns out to be $\rho^{(3)}=\frac{1}{2} I - \left(\frac{1}{2}-p \right) \tanh{\left(\beta  h \nu/2 \right)}\sigma_z$.
The work performed over the working substance by the measurement channel $\mathcal{M}^{b}$ is then given by 
$\langle \mathcal{W}\rangle=\text{tr}\left[H^{\text{C}}\left(\rho^{(3)}-\rho^{(2)}\right)\right]$. The extracted work is $\langle \mathcal{W}_{\text{ext}}\rangle=-\langle \mathcal{W}\rangle$.  By employing the map $\mathcal{M}^b$, it can be written as
\begin{equation}
    \langle \mathcal{W}_{\text{ext}}\rangle =h \nu (2p-1)\tanh \left(\frac{\beta h \nu}{2} \right).
    \label{work}
\end{equation}

\begin{figure}[t]
\centering
\includegraphics[width=8cm]{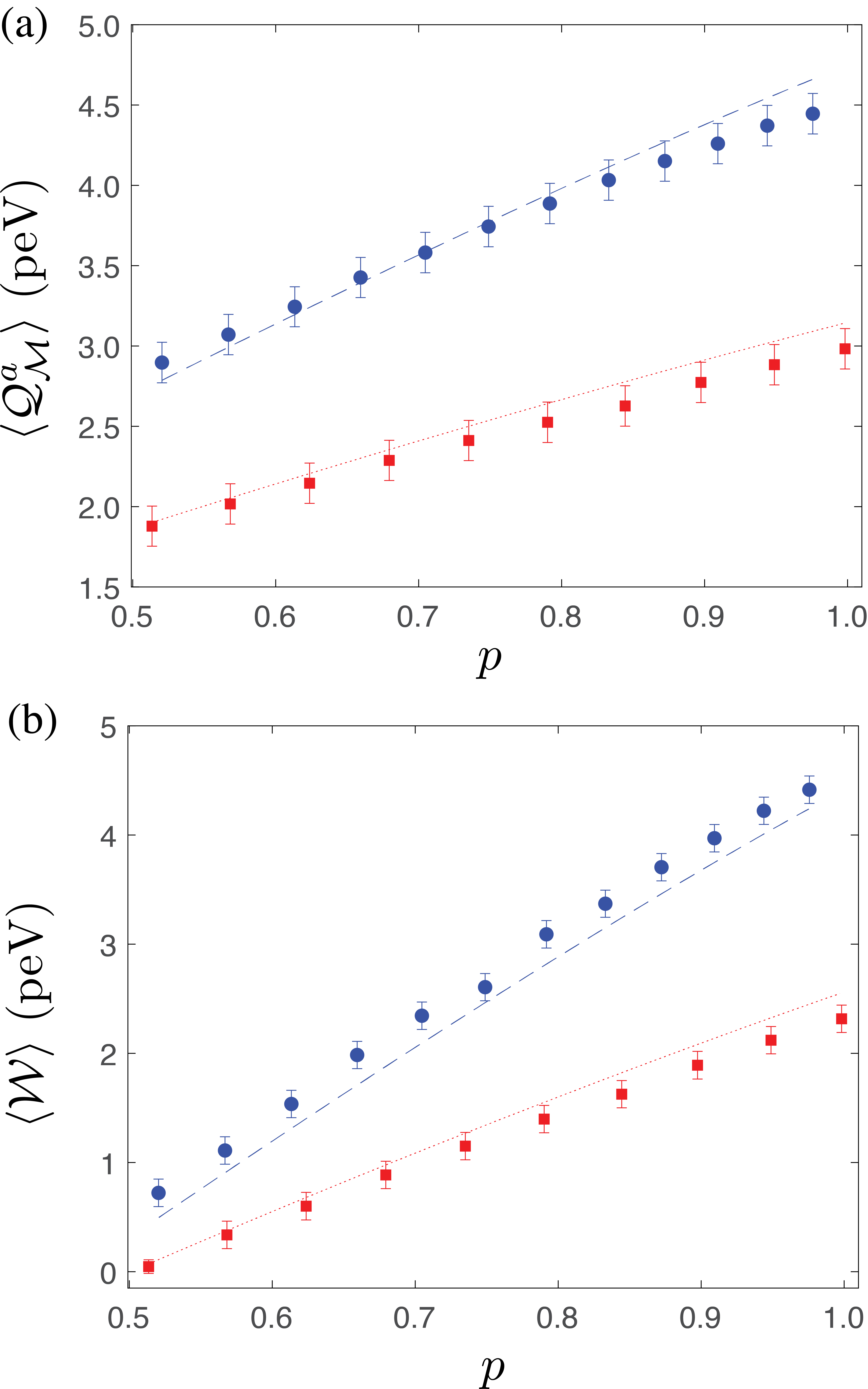}
\caption{Absorbed heat and extracted work by the generalized measurement channels. (a) Heat absorbed from the measurement channel $\mathcal{M}^{a}$ as a function of the measurement strength $p$. (b) Extracted work in the channel $\mathcal{M}^{b}$ as a function of $p$. The dashed and dotted curves represent numerical simulations of the cycle including some non-idealities. The symbols are observed experimental data with the red squares (with a theoretical dotted line) associated to the initial spin temperature $k_BT_1 = 1.88\pm 0.21$~peV while the blue circles (with a theoretical dashed line) correspond to the initial spin temperature $k_BT_2 = 2.98+\pm 0.19$~peV.}
\label{Fig3}
\end{figure}

Strokes 1, 2, and 3 can be implemented through pulse sequences described in Appendix A. The energy and von Neumann entropy variations are characterized after the sequential implementation of strokes 1, 1 followed by 2 (1-2), and 1 followed by 2 and 3 (1-2-3). In order to obtain a high-fidelity implementation, we use an optimized pulse sequence to the effective map $\mathcal{M}^b[\mathcal{M}^a(\rho^{(1)})]$ for the sequential implementation of the strokes 1-2-3.   

To quantify the performance of the generalized measurement-powered quantum engine, we use the efficiency that can be written as 
\begin{equation}
    \eta=\frac{\langle \mathcal{W}_{\text{ext}}\rangle}{\langle \mathcal{Q}^{p}\rangle}= 2-\frac{1}{p}
    \label{eq:effiency}.
\end{equation}
Depending on the measurement strength it is possible to reach the unit efficiency.
As the measurement channel takes some time to be implemented (see Appendix A), we will also characterize the extracted power per measurement cycle, defined as
\begin{equation}
    \mathcal{P}_{\text{ext}}=\frac{\langle \mathcal{W}_{\text{ext}}\rangle}{\tau_{\text{cycle}}},
\end{equation} 
with $\tau_{\text{cycle}}$ corresponding to the time to perform the pulse sequences to emulate both channels $\mathcal{M}^a$ and $\mathcal{M}^b$.  

The measurement-powered cycle was implemented varying the measurement strength for two different initial nuclear spin temperatures, corresponding to $k_BT_1 = 1.88\pm 0.21$~peV and $k_BT_2 = 2.98+\pm 0.19$~peV.

\begin{figure}[t]
\centering
\includegraphics[width=8cm]{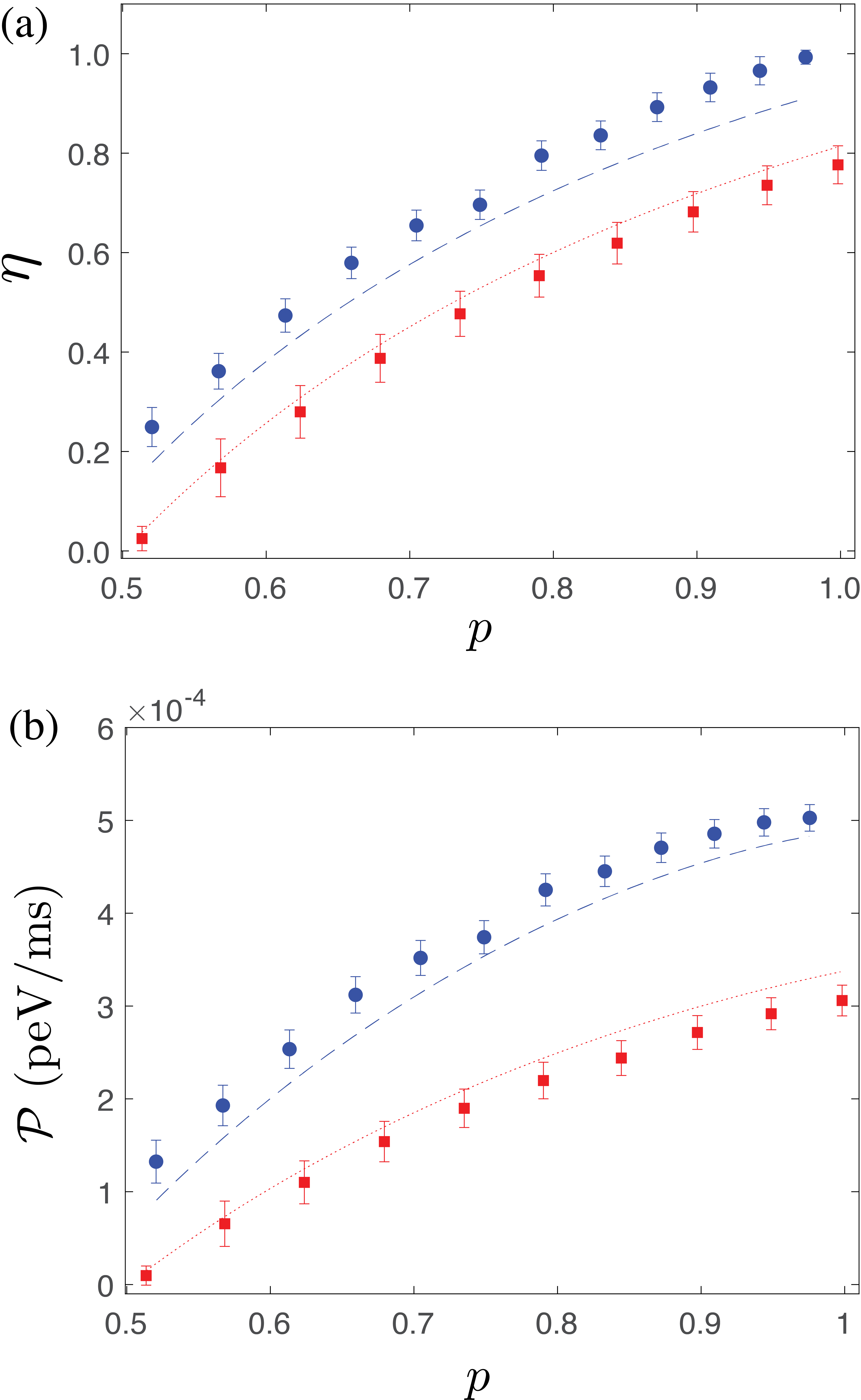}
\caption{Efficiency and extracted power for the generalized measurement-powered heat engine. (a) Efficiency of the cycle as a function measurement strength $p$. (b) Extracted power per cycle as a function of $p$.
The dashed and dotted curves represent numerical simulations of the cycle including some non-idealities. The symbols are the observed experimental data with the red squares (with a theoretical dotted line) associated to the initial spin temperature $k_BT_1 = 1.88\pm 0.21$~peV  while the blue circles (with a theoretical dashed line) correspond to the initial spin temperature $k_BT_2 = 2.98+\pm 0.19$~peV.}
\label{Fig4}
\end{figure}

In Fig. \ref{Fig2} we show the measured von Neumann entropy variation in strokes two and three, respectively. For the implantation of the measurement channel  $\mathcal{M}^a$, the entropy variation $\Delta S ^{a}>0$, highlighting the heat source nature of the meter, and it also decreases as we intensify the measurement strength (Fig. \ref{Fig2}). The choice of the measurement strength $q$ for the channel $\mathcal{M}^b$ according to Eq. (\ref{parameter}), leads to an isentropic process, that can be experimentally verified in Fig. \ref{Fig2}. Such channel feature points out the work nature of the exchanged energy with the meter.  

In Figs. \ref{Fig3}(a) and \ref{Fig3}(b) the heat absorbed in the first measurement channel, $\mathcal{M}^a$ (stroke two), and the extracted work in the second measurement channel, $\mathcal{M}^b$  (stroke three), are displayed, respectively. We experimentally confirm that, by tuning the value of the measurement strength $p$, we can adjust the amount of absorbed heat and extracted work in the cycle. A fraction of the energy absorbed by the working substance from the meter, in the form of heat in the second stroke (channel $\mathcal{M}^a$), is extracted as work in the third stroke (channel $\mathcal{M}^b$), whereas the remaining energy, $\expval{\mathcal{Q}^{\text{cold}}}=\expval{\mathcal{W}_{\text{ext}}}-\expval{\mathcal{Q}^p}$ will be dissipated to the cold sink, in the next run of the cycle (as sketched in Fig. \ref{Fig1}). 
Furthermore, in Figs. \ref{Fig4}(a) and \ref{Fig4}(b) we display the performance of the investigated engine, i.e., the efficiency and the extracted power as a function of the measurement strength $p$. As theoretically predicted, the performance of the engine increases as a function of $p$.

\section{Conclusions}

We reported on a proof-of-concept experiment of a quantum heat engine deterministically powered by generalized measurements with a nuclear spin $1/2$ playing the role of a working substance. The measurement channels are carried out with the aid of a non-observed ancillary nuclear spin $1/2$. We have evaluated the energy exchanged with the meter system in the form of heat in the first measurement and work in the second measurement channels. In the present proof-of-concept experiment, we are not concerned about how to store the work delivered to the meter system, leaving this subject to further investigations. 

From a theoretical point of view, the efficiency of the measurement-powered cycle should only depend on the measurement strength, as in Eq. (\ref{eq:effiency}). Otherwise, in a practical scenario, precise knowledge of the initial state temperature and fine control of the measurement channels parameters are required to reach the theoretical prediction and the unit efficiency. In our implementation, due to experimental non-idealities and some uncertainty about the initial spin temperature, we observed that the efficiency also depends on the initial state. In Fig. \ref{Fig4}(a), we remark that the efficiency reaches a value near to the unit for one of the initial spin temperatures considered, showing that having sufficient control over the parameters involved, it is possible to implement the measurement-powered cycle with maximum efficiency. Interestingly enough, the maximum extracted power coincides with maximum efficiency parameter tuning. Due to the intrinsically quantum nature of the measurement back-action, it is not possible to make a direct comparison with classical thermal cycles without feedback control. In this way, the generalized measurement-powered cycle studied here can be considered as a thermal device with genuine quantum features.   

The practical realization of generalized measurement channels will take, in general, a finite time that depends on its particular implementation. In our experiment, the measurement channel implementation running time is about $7.7$~ms, which is much smaller than any decoherence or thermalization time scales (of the order of seconds for our sample) in the experimental setup (see Appendix A).

We believe that our experimental investigation will motivate further developments that could unveil practical applications of measurement-powered protocols in quantum thermodynamics, along with opening new possibilities for efficiency enhancing from the combination of a system driven by a time-dependent Hamiltonian and measurements in quantum devices.

\section*{Acknowledgments}
The authors acknowledge the Brazilian funding agencies CAPES (Grants No.~88887.354951/2019-00 and No. 88887.499539/2020-00), FAPESP (Grant No.~2019/04184-5), CNPq (Grant No. 310430/2018-6 and No. 142531/2020-0), and the National Institute for Science and Technology of Quantum Information (CNPq, Grant No. INCT-IQ~465469/2014-0). The authors are grateful to the Multiuser Central Facilities (CEM-UFABC) for the experimental support.  P.R.D. acknowledges support by the Foundation for Polish Science (IRAP project, ICTQT, Contract No.~MAB/2018/5, co-financed by EU within Smart Growth Operational Programme). J.F.G.S. thanks NSFC (China) for support under Grant No.~12050410258. R.M.S. also acknowledges Ministry of Science and Technology (China), through the High-End Foreign Expert Program (Grant No.~G2021016021L).

\subsection*{Appendix A: Experimental implementation of the measurement-powered cycle}
\label{AppendixA}

\begin{figure}[h]
\centering
\includegraphics[width=8.10cm]{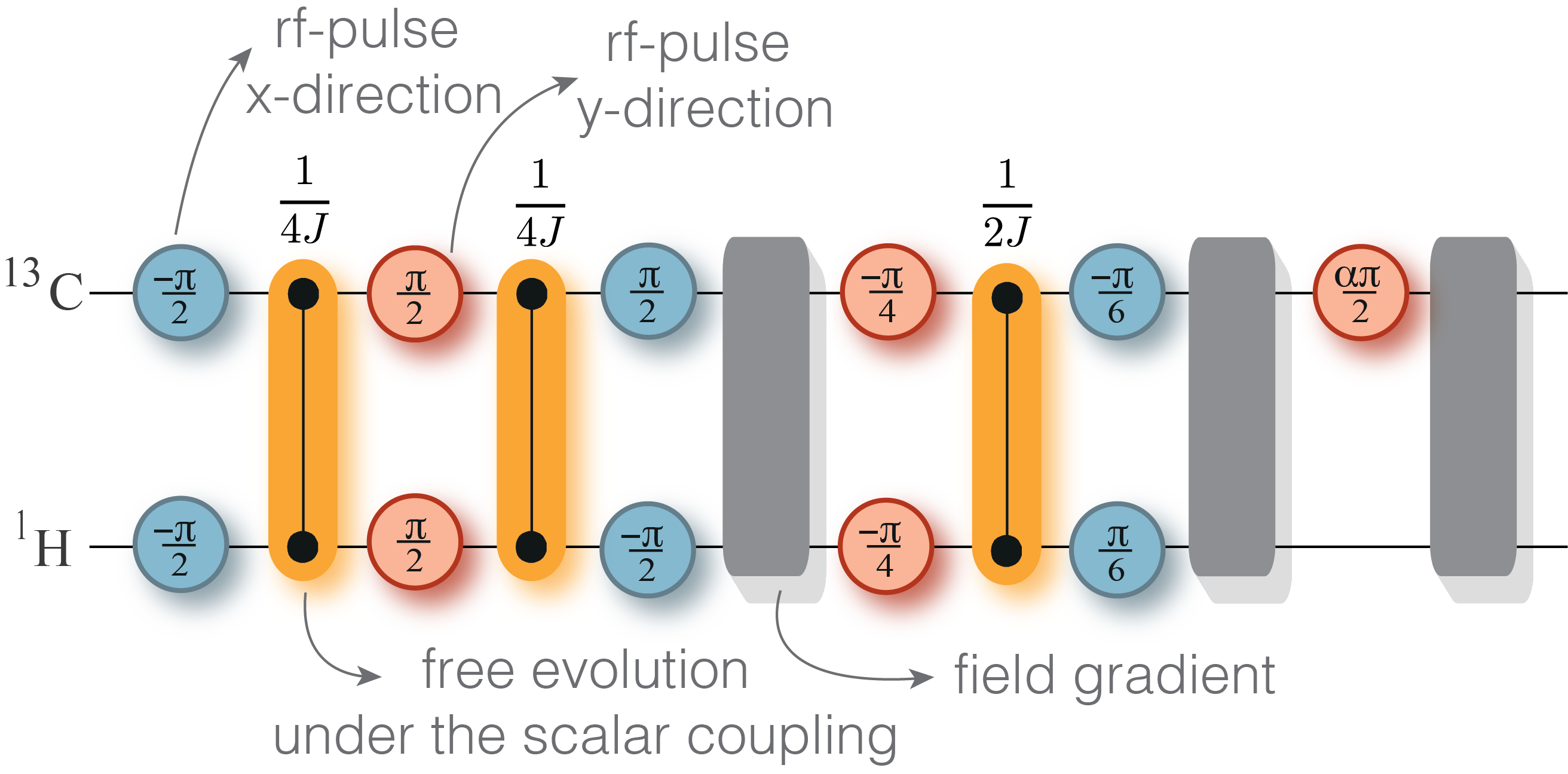}
\caption{Pulse sequence for the first stroke of the cycle - cooling. The horizontal lines represent each nuclear spin. The blue (red) circles represent $x$ ($y$) local rotations by the angles internally indicated. These rotations are produced by the transverse rf pulses on resonance or near to resonance with either $^{1}$H or $^{13}$C nuclei, with phase, amplitude, and time duration properly adjusted. The orange connections represent free evaluations under the scalar coupling of both spins, $H_J=\frac{h}{4}J\sigma_z^{\text{H}}\otimes \sigma_z^{\text{C}}$ (with $J \approx 194.65$~Hz), lasting for the time indicated above the symbol. The gray boxes stand for magnetic field gradients, with longitudinal orientations aligned to the spectrometer cylindrical symmetry axis. All the pulses were optimized to build an initial pseudo-thermal state for the $^{13}$C nuclear spin equivalent to $\rho^{(1)}=\text{exp}\left[-\beta H^{\text{C}}\right]/\mathcal{Z}$,
at cold inverse spin temperature $\beta=1/\left(k_{B}T\right)$ (which is adjusted by the angle $\alpha$). The $^{1}$H nuclear spin is prepared in a state equivalent to $\rho^{\text{H,0}} = \ket{0}\bra{0}$.} 
\label{Fig5}
\end{figure}

Effective initial states of the nuclear spins were prepared by spatial average techniques~\cite{oliveira2007,batalhao14,batalhao15}, being the $^{13}$C and $^{1}$H nuclei employed effectively as the working substance and ancillary control, respectively. The pulse sequences used to this end are depicted in Fig. \ref{Fig5}, where the $^{13}$C nuclear spin is cooled to a pseudo-equilibrium state at spin inverse temperature $\beta$, equivalent to $\rho^{(1)}=\text{exp}\left[-\beta H^{\text{C}}\right]/\mathcal{Z}$, and the ancillary  system  $^{1}$H is prepared in a state equivalent to $\rho^{\text{H},0}=\ket 0 \bra 0$.

\begin{figure}[h]
\centering
\includegraphics[width=8.10cm]{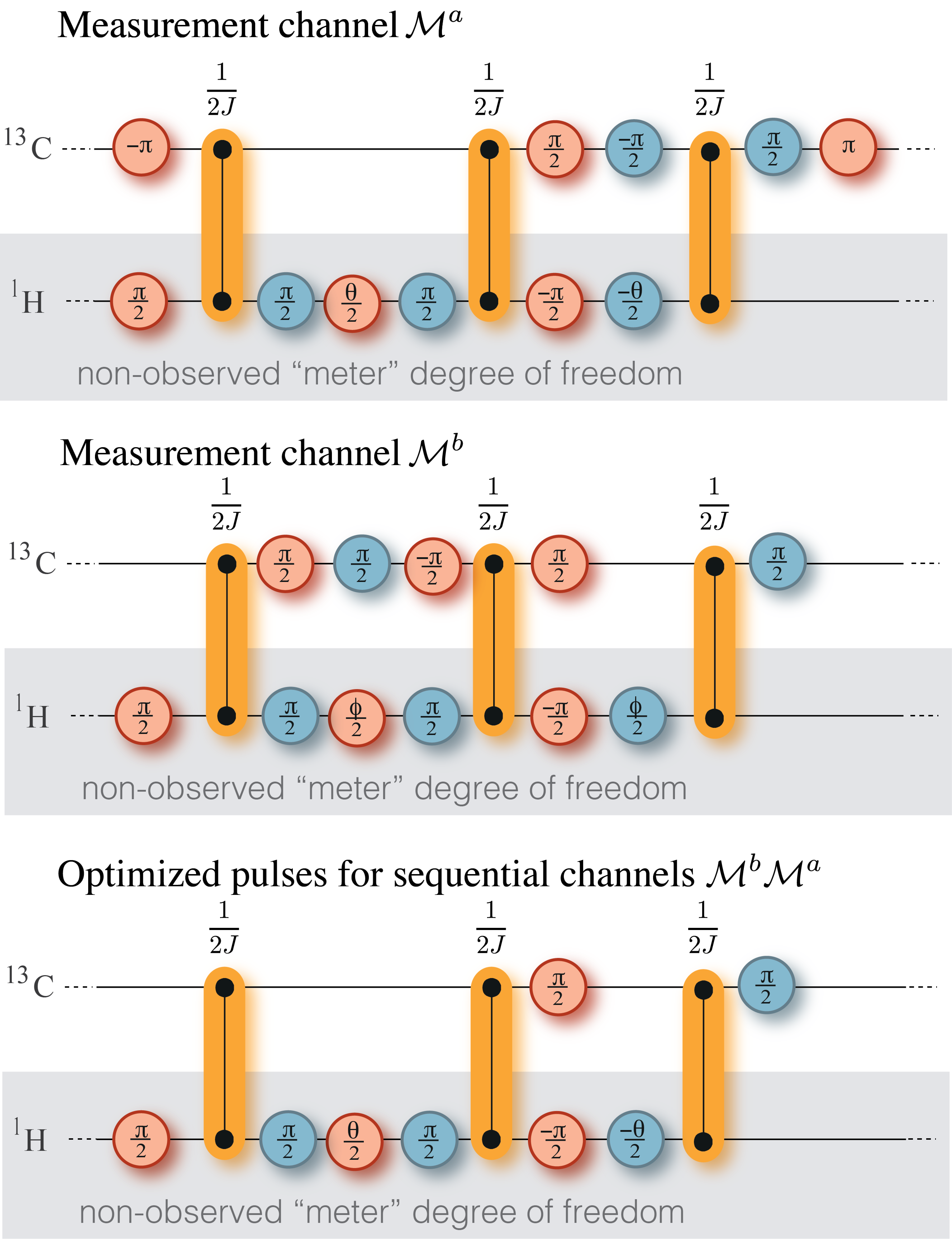}
\caption{Pulse sequences for the measurement channels implementation. The non-selective generalized measurement channel $\mathcal{M}^a$ (second stroke) is effectively performed using the sketched pulse sequence. The measurement strength parameter, $p$, is tuned by the angle $\theta \in [0,\pi/2)$ according to $\theta=\arccos{(1-2 p \Omega )}$. The channel $\mathcal{M}^b$ (third stroke) can be effectively implemented throughout the sketched pulse sequence. The strength parameter $q$ is tuned by the angle $\phi \in [0,\pi/2)$ using the relation $\phi=\arccos{(1-2q)}$. The pulse sequence for the consecutive application of $\mathcal{M}^a$ and  $\mathcal{M}^b$ is obtained by setting the strength $q$ as in Eq.~(\ref{parameter}) as a function of $p$. It was optimized, using the commutation relations for Pauli rotations to cancel redundancies, resulting in a shortened effective sequence which also minimizes errors. After the second and third strokes, the magnetization of the $^{13}$C nuclear spin is measured (determining the internal energy variation) and a QST is performed. From the last data, we obtain the variation of the von Neumann entropy. The $^{1}$H nuclear spin is not observed resulting in the desired non-selective generalized measurement channels (the CPTP maps) acting on the $^{13}$C nucleus.}  
\label{Fig6}
\end{figure}

The generalized measurement channels $\mathcal{M}^a$ and $\mathcal{M}^b$ that acts on the $^{13}$C nuclear spin are implemented through the interaction with the $^{1}$H nuclear spin which plays the role of the internal degree of the meter. At the end of the experiment, the state of the $^{1}$H nuclear spin will not be observed. So, after the pulse sequence, we only have the averaged effects of the interaction with the $^{1}$H on the $^{13}$C dynamics, leading to the desired non-selective generalized measurement map acting on the working substance ($^{13}$C). The pulse sequences that emulates the measurement channels are schematically depicted in Fig.~\ref{Fig6}. Both maps, $\mathcal{M}^a$ and $\mathcal{M}^b$, corresponding to pulse sequences in Fig.~\ref{Fig6}, lead (in general) to a non-unitary back-action to $^{13}$C nuclear spin. It is important to note that the energy exchanged (in the form of heat or work) with the meter's internal degree of freedom (emulated by the $^1$H proton) is due to the whole transformation induced by the effective CPTP maps implemented in the present proof-of-principle experiment.

For the sodium formate sample used in the NMR experiments, the spin-lattice relaxation times, measured by the inversion recovery pulse sequence, are $\mathcal{T}_1^{\text{H}}=11.67$~s and $\mathcal{T}_1^{\text{C}}=22.97$~s, for $^{1}$H on the $^{13}$C nuclear spin, respectively.  Moreover, the transverse relaxations, obtained by the Carr–Purcell–Meiboom–Gill
pulse sequence, have characteristic times $\mathcal{T}_2^{\text{H}}=1.31$~s and $\mathcal{T}_2^{\text{C}}=2.57$~s, for $^{1}$H on the $^{13}$C nuclear spin, respectively. We note that the measurement channels implementation as depicted in the pulse sequences Fig.~\ref{Fig6} have a running time of about $7.7$~ms which is much smaller than the spin-lattice and transverse relaxation time.

\subsection*{Appendix B: Error analysis}

The main sources of experimental errors are small uncontrolled variations in the transverse rf-field intensities, non-idealities in its time modulation, and tiny inhomogeneities in the longitudinal static field as well as in the gradient pulses. All pulses in the experiment were optimized to minimize such errors. 

The error bars shown in Figs.~\ref{Fig2}, \ref{Fig3}(a) and \ref{Fig3}(b), and \ref{Fig4}(a) and \ref{Fig4}(b) were Monte Carlo estimated, sampling deviations of the nuclear magnetization or quantum state tomography (QST) data with a Gaussian distribution having widths determined by the variances corresponding to such data. The variances of the magnetization measurements and QST data are obtained by preparing the same pseudo thermal state $100$ times and comparing it with the theoretical expectation. These variances include random and systematic errors in both initial state preparation and data acquisition.


\begin{thebibliography}{70}



\bibitem{Ce2001book}Y. A. Cengel and M. A. Boles, Thermodynamics. An Engineering Approach-(McGraw-Hill, New York, 2001).

\bibitem{Myers2022} Nathan M. Myers, Obinna Abah, and Sebastian Deffner, Quantum thermodynamic devices: from theoretical proposals to experimental reality, \href{https://doi.org/10.1116/5.0083192}{AVS
Quantum Sci. \textbf{4}, 027101 (2022).}

\bibitem{campisi2011colloquium} M. Campisi,  P. H{\"a}nggi, P. Talkner, Colloquium: Quantum fluctuation relations: Foundations and applications, \href{https://doi.org/10.1103/RevModPhys.83.771}{Rev. Mod. Phys. \textbf{83}, 771 (2011)}.

\bibitem{campisi14} M. Campisi, Fluctuation relation for quantum heat engines and refrigerators, \href{https://doi.org/10.1088/1751-8113/47/24/245001}{J. Phys. A: Math. Theor. \textbf{47}, 245001 (2014)}.


\bibitem{batalhao14} T. B. Batalh\~{a}o,  A. M. Souza, L. Mazzola, R. Auccaise, R. S. Sarthour, I. S. Oliveira, J. Goold, G. De Chiara, M. Paternostro, and R. M. Serra, Experimental Reconstruction of Work Distribution and Study of Fluctuation Relations in a Closed Quantum System, \href{https://doi.org/10.1103/PhysRevLett.113.140601}{Phys. Rev. Lett. \textbf{113}, 140601 (2014)}.

\bibitem{elouard15} C. Elouard, M. Richard, and A. Auff\`{e}ves, Reversible work extraction in a hybrid opto-mechanical system, \href{https://doi.org/10.1088/1367-2630/17/5/055018}{New J. Phys. \textbf{17}, 055018 (2015)}.

\bibitem{campisi15} M. Campisi, J. Pekola, and R. Fazio, Nonequilibrium fluctuations in quantum heat engines: theory, example, and possible solid state experiments, \href{https://doi.org/10.1088/1367-2630/17/3/035012}{New J. Phys. 17, 035012 (2015)}.

\bibitem{campisi16} M. Campisi and R. Fazio, The power of a critical heat engine, \href{https://doi: 10.1038/ncomms11895 (2016)}{Nat. Commun. \textbf{7}, 11895 (2016)}.

\bibitem{dechant15} A. Dechant, N. Kiesel, and E. Lutz, All-Optical Nanomechanical Heat Engine, \href{https://doi.org/10.1103/PhysRevLett.114.183602}{Phys. Rev. Lett. \textbf{114}, 183602 (2015)}.

%
\bibitem{Peterson2018}J. P. S. Peterson, T. B. Batalhão, M. Herrera, A. M.
Souza, R. S. Sarthour, I. S. Oliveira, R. M. Serra, Experimental Characterization of a Spin Quantum Heat Engine, \href{https://doi.org/10.1103/PhysRevLett.123.240601}{Phys. Rev. Lett. \textbf{123}, 240601 (2019)}.

\bibitem{Klatzow2019} J. Klatzow, J. N. Becker, P. M. Ledingham, C. Weinzetl,
K. T. Kaczmarek, D. J. Saunders, J. Nunn, I. A. Walmsley, R. Uzdin, and E. Poem, Experimental Demonstration of Quantum Effects in the Operation of Microscopic Heat
Engines, \href{https://doi.org/10.1103/PhysRevLett.122.110601}{Phys. Rev. Lett. \textbf{122}, 110601 (2019)}.

\bibitem{Zagoskin2012} A. M. Zagoskin, S. Savel’ev, Franco Nori, and F. V. Kusmartsev, Squeezing as the source of inefficiency in the quantum Otto cycle, \href{https://doi.org/10.1103/PhysRevB.86.014501}{Phys. Rev. B \textbf{86}, 014501 (2012)}.

\bibitem{Altintas2015} F. Altintas and Ö. E. Müstecaplıoğlu, General formalism of local thermodynamics with an example: Quantum Otto engine with a spin-1/2 coupled to an arbitrary spin, \href{https://doi.org/10.1103/PhysRevE.92.022142}{Phys. Rev. E \textbf{92}, 022142 (2015)}.

\bibitem{Campisi2016} M. Campisi and R. Fazio, Dissipation, correlation and lags in heat engines, \href{https://doi.org/10.1088/1751-8113/49/34/345002}{J. Phys. A: Math. Theor. \textbf{49}, 345002
(2016)}.

\bibitem{Rossnagel2016} J. Roßnagel, S. T. Dawkins, K. N. Tolazzi, O. Abah, E. Lutz, F. Schmidt-Kaler, and K. Singer, A single-atom heat engine, \href{https://www.science.org/doi/10.1126/science.aad6320}{Science \textbf{352}, 325 (2016)}.

\bibitem{Barontini2019} G. Barontini and M. Paternostro, Ultra-cold single-atom quantum heat engines, \href{https://doi.org/10.1088/1367-2630/ab2684}{New J. Phys. \textbf{21} 063019 (2019)}.

\bibitem{Bouton2021} Q. Bouton, J. Nettersheim, S. Burgardt, D. Adam, E. Lutz, and  A. Widera, A quantum heat engine driven by atomic collisions, \href{https://doi.org/10.1038/s41467-021-22222-z}{Nat Commun \textbf{12}, 2063 (2021)}.

\bibitem{esposito2009nonequilibrium} M. Esposito, U. Harbola, and S. Mukamel, Nonequilibrium fluctuations, fluctuation theorems, and counting statistics in quantum systems, \href{https://doi.org/10.1103/RevModPhys.81.1665}{Reviews of modern physics \textbf{81}, 1665 (2009)}.

\bibitem{quan07} H. T. Quan, Y. xi Liu, C. P. Sun, and F. Nori, Quantum thermodynamic cycles and quantum heat engines, \href{https://doi.org/10.1103/PhysRevE.79.041129}{Phys. Rev. E 76, 031105 (2007)}.


\bibitem{Kosloff2013} R. Kosloff, Quantum thermodynamics: A dynamical
viewpoint, \href{https://doi.org/10.3390/e15062100}{Entropy \textbf{15}, 2100 (2013)}.

\bibitem{Alicki2019} R. Alicki and R. Kosloff, \textquotedblleft Introduction
to quantum thermodynamics: History and prospects\textquotedblright , in Thermodynamics in the Quantum Regime, edited by  F. Binder et al., (Springer,Cham, 2019), pp. 1--33, and references therein.

\bibitem{Vinjanampathy2016}S. Vinjanampathy and J. Anders, Quantum
thermodynamics, \href{https://doi.org/10.1080/00107514.2016.1201896}{Contemp. Phys. \textbf{57}, 545 (2016)}.

\bibitem{Timpanaro2019} A. M. Timpanaro, G. Guarnieri, J.Goold, and
G. T. Landi, Thermodynamic Uncertainty Relations from Exchange Fluctuation Theorems, \href{https://doi.org/10.1103/PhysRevLett.123.090604}{Phys. Rev. Lett. \textbf{123}, 090604 (2019)}.

\bibitem{Lee2021}S. Lee, M. Ha, and H. Jeong, Quantumness and thermodynamic
uncertainty relation of the finite-time Otto cycle, \href{https://doi.org/10.1103/PhysRevE.103.022136}{Phys. Rev. E \textbf{103},
022136 (2021)}.

\bibitem{Sacchi2021} M. F. Sacchi, Thermodynamic uncertainty relations
for bosonic Otto engines, \href{https://doi.org/10.1103/PhysRevE.103.012111}{Phys. Rev. E \textbf{103}, 012111 (2021)}.


\bibitem{berg2018}J. Åberg, Fully Quantum Fluctuation Theorems,
\href{https://doi.org/10.1103/PhysRevX.8.011019}{Phys. Rev. X \textbf{8}, 011019 (2018)}.

\bibitem{micadei2020} K. Micadei, G. T. Landi, and E. Lutz, Quantum Fluctuation
Theorems beyond Two-Point Measurements, \href{https://doi.org/10.1103/PhysRevLett.124.090602}{Phys. Rev. Lett. \textbf{124},
090602 (2020)}.



\bibitem{Denzler2021} T. Denzler, J. F. G. Santos, E. Lutz, and R.
M. Serra, Nonequilibrium fluctuations of a quantum heat engine, \href{https://arxiv.org/abs/2104.13427}{arXiv:2104.13427}.




\bibitem{Groot1984}S. R. de Groot and P. Mazur, Non-Equilibrium Thermodynamics,
(Dover Publications, Inc., New York, 1984).

\bibitem{Lebon2008}G. Lebon, D. Jou, and J. Casas-Vázques, Understading
Non-Equilibrium Thermodynamics: Foundations, Applications, Frontiers,
(Springer-Verlag, Berlin, 2008).

\bibitem{batalhao15}  T. B. Batalh\~{a}o, A. M. Souza, R. S. Sarthour, I. S. Oliveira, M. Paternostro, E. Lutz, and R. M. Serra,  Irreversibility and the Arrow of Time in a Quenched Quantum System, \href{https://doi.org/10.1103/PhysRevLett.115.190601}{Phys. Rev. Lett. \textbf{115}, 190601 (2015)}.

\bibitem{camati16} P. A. Camati, J. P. S. Peterson, T. B. Batalh\~{a}o, K. Micadei, A. M. Souza, R. S. Sarthour, I. S. Oliveira, and R. M. Serra, Experimental Rectification of Entropy Production by Maxwell's Demon in a Quantum System, \href{https://doi.org/10.1103/PhysRevLett.117.240502}{ Phys. Rev. Lett. \textbf{117}, 240502 (2016)}.

\bibitem{Henao2018} C. I. Henao and R. M. Serra, Role of quantum coherence in the thermodynamics of energy transfer, \href{https://doi.org/10.1103/PhysRevE.97.062105}{Phys. Rev. E \textbf{97}, 062105 (2018)}. 

\bibitem{Micadei2019} K. Micadei, J. P. S. Peterson, A. M. Souza, R. S. Sarthour, I. S. Oliveira, G. T. Landi, T. B. Batalhão, R. M. Serra, and E. Lutz, Reversing the direction of heat flow using quantum correlations, \href{ https://doi.org/10.1038/s41467-019-10333-7}{Nat. Comm. \textbf{10}, 2456 (2019)}.

\bibitem{Camati2019}P. A. Camati, J. F. G. Santos, and R. M. Serra,
Coherence effects in the performance of the quantum Otto heat engine,
\href{https://doi.org/10.1103/PhysRevA.99.062103}{Phys. Rev. A \textbf{99}, 062103 (2019)}.

\bibitem{Francica2019} G. Francica, J. Goold, and F. Plastina, The
role of coherence in the non-equilibrium thermodynamics of quantum
systems, \href{https://doi.org/10.1103/PhysRevE.99.042105}{Phys. Rev. E \textbf{99}, 042105 (2019)}.

\bibitem{Santos2019}J. P. Santos, L. C. C\'{e}leri, G. T. Landi, and
M. Paternostro, The role of quantum coherence in non-equilibrium entropy
production, \href{https://doi.org/10.1038/s41534-019-0138-y}{npj Quantum Inf. 5, 23 (2019)}.


\bibitem{Szilard1929} L. Szilard, über die Entropieverminderung in einem thermodynamischen System bei Eingriffen intelligenter Wesen, \href{https://doi.org/10.1007/BF01341281}{Zeitschrift fur Physik \textbf{53},
840 (1929)}.


\bibitem{Quan2006} H.T. Quan, Y.D. Wang, Y.X. Liu, C.P. Sun, and F.
Nori, Maxwell’s Demon Assisted Thermodynamic Cycle in Superconducting Quantum Circuits, \href{https://doi.org/10.1103/PhysRevLett.97.180402}{Phys. Rev. Lett. 97, 180402 (2006)}.

\bibitem{Leff1990} Maxwell's
Demon: Entropy, Information, Computation, Computing, edited by H.S. Leff, and A.F. Rex (Princeton University
Press, Princeton, 1990).


\bibitem{Elouard2017} C.Elouard, D. Herrera-Mart\'{i}, B. Huard, and A.Auff\`{e}ves,
Extracting Work from Quantum Measurement in Maxwell's
Demon Engines, \href{https://doi.org/10.1103/PhysRevLett.118.260603}{Phys. Rev. Lett. \textbf{118}, 260603 (2017)}.

\bibitem{Elouard2018}C. Elouard and A.N. Jordan, Efficient Quantum
Measurement Engine, \href{https://doi.org/10.1103/PhysRevLett.120.260601}{Phys. Rev. Lett. \textbf{120}, 260601 (2018)}.

\bibitem{Talkner2017} J. Yi, P. Talkner, and Y. W. Kim, Single-temperature
quantum engine without feedback control, \href{https://10.1103/physreve.96.022108}{Phys. Rev. E 96, 022108 (2017)}.

\bibitem{Elouard2017-1} C. Elouard, D. Herrera-Mart\'{i}, M. Clusel, and
A. Auff\`{e}ves, The role of quantum measurement in stochastic thermodynamics,
\href{https://doi.org/10.1038/s41534-017-0008-4}{npj Quantum Inf. \textbf{3}, 9 (2017)}.

\bibitem{Brandner2015} K. Brandner, M. Bauer, M. T. Schmid and U.
Seifert, Coherence-enhanced efficiency of feedback-driven quantum
engines, \href{https://doi.org/10.1088/1367-2630/17/6/065006}{New J. Phys. \textbf{17}, 065006 (2015)}.

\bibitem{Lin2021} Z. Lin, S. Su, J. Chen, J. Chen, and  J. F. G. Santos, Suppressing coherence effects in quantum-measurement-based engines, \href{https://doi.org/10.1103/PhysRevA.104.062210}{Phys. Rev. A \textbf{104}, 062210 (2021)}.

\bibitem{Su2021} S. Su, Z. Lin, and J. Chen, Thermal divergences of quantum measurement engine, \href{https://arxiv.org/abs/2109.10796}{arXiv:2109.10796}.

\bibitem{Bresque2021} L. Bresque, P. A. Camati, S. Rogers, K. Murch, A. N. Jordan, and A. Auffèves, Two-Qubit Engine Fueled by Entanglement and Local Measurements, \href{https://doi.org/10.1103/PhysRevLett.126.120605}{Phys. Rev. Lett. \textbf{126}, 120605 (2021)}.

\bibitem{Campisi2019} L. Buffoni, A. Solfanelli, P. Verrucchi, A. Cuccoli, and M. Campisi, Quantum Measurement Cooling, \href{https://doi.org/10.1103/PhysRevLett.122.070603}{Phys. Rev. Lett. \textbf{122}, 070603 (2019)}.

\bibitem{Ding2018} X. Ding, J. Yi, Y. W. Kim, and P. Talkner, Measurement-driven single temperature engine, \href{https://doi.org/10.1103/PhysRevE.98.042122}{ Phys. Rev. E \textbf{98}, 042122 (2018)}. 

\bibitem{Jordan2020} A. Jordan, C. Elouard, and A. Auffèves, Quantum measurement engines and their relevance for quantum interpretations, \href{https://doi.org/10.1007/s40509-019-00217-2}{Quantum Stud: Math. Found. \textbf{7}, 203 (2020)}. 

\bibitem{Mohammady2017} M. H. Mohammady and J. Anders, A quantum Szilard engine without heat from a thermal reservoir, \href{https://doi.org/10.1088/1367-2630/aa8ba1}{New J. Phys. \textbf{19}, 113026 (2017)}.

\bibitem{Campisi2017} M. Campisi, J. Pekola, and R. Fazio, Feedback-controlled heat transport in quantum devices: theory and solid-state experimental proposal, \href{https://doi.org/10.1088/1367-2630/aa6acb}{New J. Phys. \textbf{19}, 053027 (2017)}.

\bibitem{chand2017single} S. Chand and A. Biswas, Single-ion quantum Otto engine with always-on bath interaction, \href{https://doi.org/10.1209/0295-5075/118/60003}{Europhys. Lett. \textbf{118}, 60003 (2017)}.


\bibitem{chand2017measurement} S. Chand and A. Biswas, Measurement-induced operation of two-ion quantum heat machines, \href{https://doi.org/10.1103/PhysRevE.95.032111}{Phys. Rev. E \textbf{95}, 032111 (2017)}.

\bibitem{chand2018critical} S. Chand and A. Biswas, Critical-point behavior of a measurement-based quantum heat engine, \href{https://doi.org/10.1103/PhysRevE.98.052147}{Phys. Rev. E \textbf{98}, 052147 (2018)}.


\bibitem{anka2021measurement} M. F. Anka, T. R. de Oliveira, and D. Jonathan, Measurement-based quantum heat engine in a multilevel system, \href{https://doi.org/10.1103/PhysRevE.104.054128}{Phys. Rev. E \textbf{104}, 054128 (2021)}.

\bibitem{alam2022two} M. S. Alam, and B. P. Venkatesh, Two-stroke quantum measurement heat engine, \href{
https://doi.org/10.48550/arXiv.2201.06303}{	arXiv:2201.06303 (2022)}.


\bibitem{mancino18} L. Mancino, M. Sbroscia, E. Roccia, et al, The entropic cost of quantum generalized measurements. \href{https://doi.org/10.1038/s41534-018-0069-z}{npj Quantum Inf \textbf{4}, 20 (2018)}.

\bibitem{jacobs09} K. Jacobs, The second law of thermodynamics and quantum feedback control: Maxwell's demon with weak measurements, \href{https://doi.org/10.1103/PhysRevA.80.012322}{Phys. Rev. A \textbf{80}, 012322 (2009)}.

\bibitem{pati2020} A. K. Pati, C. Mukhopadhyay, S. Chakraborty, and S. Ghosh, Quantum precision thermometry with weak measurements, \href{https://doi.org/10.1103/PhysRevA.102.012204}{Phys. Rev. A \textbf{102}, 012204 (2020)}.


\bibitem{alonso16} J. J. Alonso, E. Lutz, and A. Romito, Thermodynamics of Weakly Measured Quantum Systems, \href{https://doi.org/10.1103/PhysRevLett.116.080403}{Phys. Rev. Lett. \textbf{116}, 080403 (2016)}.

\bibitem{oreshkov05}  O. Oreshkov, T. Brun, Weak Measurements are Universal, \href{https://doi.org/10.1103/PhysRevLett.95.110409}{Phys. Rev. Lett. \textbf{95}, 1104909 (2005)}.

\bibitem{Dieguez18} P. R. Dieguez, and R.M. Angelo, Information-reality complementarity: The role of measurements and quantum reference frames, \href{https://doi.org/10.1103/PhysRevA.97.022107}{Phys. Rev. A {\bf97}, 022107 (2018)}.

\bibitem{pan16} Y. Pan, J. Zhang, E. Cohen, et al, Weak-to-strong transition of quantum measurement in a trapped-ion system, \href{https://doi.org/10.1038/s41567-020-0973-y}{Nat. Phys. \textbf{16}, 1206–1210 (2020)}. 

\bibitem{naghiloo2020heat} M. Naghiloo,  D. Tan, P. M. Harrington, J. J. Alonso, E. Lutz, A. Romito, and K. W. Murch, Heat and Work Along Individual Trajectories of a Quantum Bit, \href{https://doi.org/10.1103/PhysRevLett.124.110604}{Phys. Rev. Lett. \textbf{124}, 110604 (2020)}. 

\bibitem{naghiloo2018information} M. Naghiloo, J. J. Alonso, A. Romito, E. Lutz, and K. W. Murch, Information Gain and Loss for a Quantum Maxwell's Demon, 
\href{https://doi.org/10.1103/PhysRevLett.121.030604}{Phys. Rev. Lett. \textbf{121}, 030604  (2018)}.

\bibitem{hernandez2022autonomous}
S. Hernández-Gómez, S. Gherardini, N. Staudenmaier, F. Poggiali, M. Campisi, A. Trombettoni, F.S. Cataliotti, P. Cappellaro, and N. Fabbri, Autonomous Dissipative Maxwell's Demon in a Diamond Spin Qutrit, \href{https://doi.org/10.1103/PRXQuantum.3.020329}{PRX Quantum \textbf{3}, 020329 (2022)}
.

\bibitem{behzadi2020quantum} N.  Behzadi,  Quantum  engine  based  on  general  measurements, \href{https://10.1088/1751-8121/abca74}{J. Phys. A: Math. and Theor. \textbf{54}, 015304 (2021)}.


\bibitem{oliveira2007} I. S. Oliveira, T. J. Bonagamba, R. S. Sarthour, J. C. C. Freitas, R. R. de Azevedo, \textit{NMR Quantum Information
Processing} \href{https://doi.org/10.1016/B978-0-444-52782-0.X5000-3}{(Elsevier, Amsterdam, 2007)}.

\bibitem{Micadei2021} K. Micadei, J. P. S. Peterson, A. M. Souza, R. S. Sarthour, I. S. Oliveira, G. T. Landi, R. M. Serra, and E. Lutz, Experimental Validation of Fully Quantum Fluctuation Theorems Using Dynamic Bayesian Networks, \href{https://doi.org/10.1103/PhysRevLett.127.180603}{Phys. Rev. Lett. \textbf{127}, 180603 (2021)}.

\bibitem{Kim2012} Y-S. Kim, J-C. Lee, O. Kwon and Y-H. Kim, \href{https://doi.org/10.1038/nphys2178}{Nat. Phys, \textbf{8}, 117 (2012)}.



\end{thebibliography}
\end{document}